# Magnetic ordering in mackinawite (tetragonal FeS): evidence for strong itinerant spin fluctuations

**Kideok D. Kwon**[1*], **Keith Refson**[2], **Sharon Bone**[3], **Ruimin Qiao**[4], **Wan-li Yang**[4], **Zhi Liu**[4], and **Garrison Sposito**[1,3]

[1]Geochemistry Department, Earth Sciences Division, Lawrence Berkeley National Laboratory, Berkeley, California 94720, USA

[2]STFC Rutherford Appleton Laboratory, Didcot, Oxfordshire OX11 0QX, UK

[3]Department of Environmental Science, Policy and Management, University of California, Berkeley, California 94720, USA

[4]Advanced Light Source, Lawrence Berkeley National Laboratory, Berkeley, California 94720, USA

*Author to whom correspondence should be addressed: kkwon@lbl.gov

**Abstract**

Mackinawite is a naturally-occurring layer type FeS mineral well-known to be important in biogeochemical cycles and, more recently, to the development of microbial fuel cells. Conflicting results have been published as to the magnetic properties of this mineral, with Mössbauer spectroscopy indicating no magnetic ordering and density functional theory predicting an antiferromagnetic ground state, similar to the Fe-based high-temperature superconductors with which it is isostructural and for which it is known that magnetism is suppressed by strong itinerant spin fluctuations. We investigated this latter possibility for mackinawite using photoemission spectroscopy (PES), near-edge X-ray absorption fine structure spectroscopy (XAS), and density functional theory (DFT) computations. Our Fe 3$s$ core-level PES spectrum of mackinawite showed a clear exchange-energy splitting (2.9 eV) consistent with a 1 $\mu_B$ magnetic moment on the Fe ions, while the Fe $L$-edge XAS spectrum indicated rather delocalized Fe 3$d$ electrons in mackinawite similar to those in Fe metal. DFT computations demonstrated that the ground state of mackinawite is single-stripe antiferromagnetic, with an Fe magnetic moment (2.7 $\mu_B$) that is significantly larger than the experimental estimate and has a strong dependence on the S height and lattice parameters. All of these trends signal the existence of strong itinerant spin fluctuations. If strong spin fluctuations are the mediators of electron pairing in these Fe-based superconductors, we conjecture that mackinawite may be one of the simplest Fe-based superconductors.

# I. INTRODUCTION

Mackinawite (tetragonal FeS)[1] is a layer type mineral comprising Fe(II) arranged in tetrahedral coordination with S(-II) on a square lattice to form edge-sharing $FeS_4$ tetrahedral sheets stacked along the *c*-axis and stabilized via van der Waals forces (Fig. 1). Its lattice parameters (*P4/nmm* space group) are $a = b = 3.674$ Å and $c = 5.033$ Å.[2] In nature, this mineral is typically the first solid phase to be precipitated during iron sulfide mineralization mediated by sulfate-reducing bacteria.[3-5] Over time, extending up to two years in anoxic sediments, it transforms into greigite ($Fe_3S_4$), pyrrhotite [hexagonal or monoclinic $Fe_{(1-x)}S$, $0 < x < 0.2$], pyrite ($FeS_2$), or, depending on redox conditions, elemental S and Fe.[6-9] Mackinawite is known to influence the mobility and bioavailability of environmentally-important trace elements, notably through processes involving either sorption[10, 11] or oxidative dissolution,[12-14] and it figures in the well-known iron-sulfur hypothesis on the origin of life.[15] Very recently, Nakamura et al.[16] demonstrated that cell-enmeshed assemblies of mackinawite produced by the Fe-reducing bacterium, *Shewanella loihica* PV-4, function efficiently as electron-transfer conduits and, therefore, have potential for use as bioanode materials in microbial fuel cells.

The electronic properties of mackinawite are not as clearly established as its structural parameters. An electrical conductivity for mackinawite is not available, but the mineral has been considered to be metallic (i.e., highly delocalized Fe *3d* electrons) based on its short Fe-Fe distance (2.60 Å), which is close to that in elemental iron (2.48 Å for *α*-iron). Mössbauer spectroscopy indicates that mackinawite has no magnetic ordering down to 4.2 K.[17, 18] The spectra exhibit one singlet without quadrupole splitting and hyperfine coupling (isomer shift 0.2 mms$^{-1}$ with respect to elemental iron) between room temperature and 4.2 K, and application of high magnetic fields (3 T) does not reveal an internal field contribution from electrons. This temperature-independent Mössbauer behavior along with the short Fe-Fe distance lead one to hypothesize that mackinawite is only Pauli paramagnetic.[19] Recent density functional theory structural optimizations of mackinawite, however, have come to opposing conclusions about the magnetic moment on the Fe ions. Devey et al.[20] concluded that the stable ground state of the mineral is non-magnetic (NM), whereas Subedi et al.[21] proposed that ground state mackinawite with the experimental crystal structure should have a substantial magnetic moment on its Fe ions.

Mackinawite in fact has magnetic characteristics that are remarkably similar to those of the Fe-based high-temperature superconductors with which it is isostructural (e.g., FeSe$_{0.88}$ with $T_c$ = 12 K and LaFeAsO$_{0.89}$F$_{0.11}$ with $T_c$ = 26 K).[22, 23] (The layer type FeAs compounds have La, O, and F atoms interposed as spacers between edge-sharing FeAs$_4$ tetrahedral sheets.) These compounds exhibit the same temperature-independent Mössbauer spectral features down to 4.2 K as found for mackinawite and they show no internal magnetic field under applied magnetic fields up to 7 T.[24-26] In the case of undoped LaFeAsO, which does not undergo a superconducting transition, a small magnetic moment on the Fe ions (0.36 $\mu_B$) has been observed,[24] whereas in FeSe, a magnetic moment has not been observed regardless of the stoichiometry.[25] By contrast, density functional theory (DFT) studies predict that these layer-type Fe-based compounds, either doped or undoped, have a magnetic ground state, in particular, one with antiferromagnetic (AFM) single-stripe ordering (Fig. 2), with a large magnetic moment on the Fe ions (2 $\mu_B$).[27-31] This conflict between Mössbauer spectroscopy and DFT calculations can be resolved if the existence of strong itinerant spin fluctuations is hypothesized: the ground state is indeed magnetic with AFM ordering, but spin fluctuations suppress this magnetism, and, therefore, Mössbauer spectroscopy detects no magnetic ordering and conventional DFT cannot capture the essential physics.[32, 33] These spin fluctuations also are believed to be associated with electron pairing as important mediators in the Fe-based superconductors.[34, 35]

Direct experimental evidence for the existence of strong itinerant spin fluctuations in layer type Fe-based superconductors has been obtained by measuring the Fe 3$s$ core-level photoemission spectrum.[36, 37] Multiplet splitting occurs in the spectrum mainly by the exchange interaction of core 3$s$ electrons with the net spin of 3$d$ electrons.[38-40] If Fe ions have no magnetic moment, no splitting is observed.[41] Itinerantly-magnetic Fe materials (e.g., CeFeAsF$_{0.11}$O$_{0.89}$ and NbFe$_2$) that are non-magnetic according to Mössbauer spectroscopy exhibit a large Fe 3$s$ core-level splitting, corresponding to an Fe magnetic moment of about 1 $\mu_B$.[36, 42-45]

Besides this spectroscopic evidence, the existence of spin fluctuations is signaled by a large DFT over-estimation of the Fe magnetic moment as compared to experiment.[32, 33] DFT in fact typically under-estimates the Fe magnetic moment in strongly-correlated materials (e.g., α-Fe$_2$O$_3$). A strong dependence of the Fe magnetic moment on the $z$-coordinate of the ligand (e.g.,

S or As) along the *c* axis is also commonly observed in layer type Fe-based superconductors.[33, 46] Subedi et al.[21] reported a significant increase in the Fe magnetic moment with elevation of the S-ligands in mackinawite above their location as found by conventional DFT optimization, thus suggesting the presence of strong itinerant spin fluctuations akin to those in Fe-based superconductors.

In this paper, we report conclusive spectroscopic evidence and DFT calculations supporting the existence of strong itinerant spin fluctuations in mackinawite. We examine the Fe 3*s* core-level photoemission spectrum and Fe *L*-edge spectra of the mineral collected by using photoemission spectroscopy (PES) and X-ray absorption spectroscopy (XAS). Total energy, structural parameters, and Fe magnetic moment are compared among NM, AFM, and ferromagnetic (FM) ordering in DFT calculations with the generalized gradient approximation (GGA).

## II. METHODOLOGY

### A. Experimental measurements

Mackinawite samples were synthesized under strictly anoxic conditions in a glove box consisting of a mixture of $H_2$ (3.7 %) and $N_2$, by mixing 450 μL of 0.5 M Fe(II) solution [$(NH_4)_2Fe(SO_4)_2·6H_2O$] with 450 μL of 0.42 M S(-II) solution [$Na_2S·9H_2O$] in 20 mL of 10 mM HEPES buffer solution (pH = 6.8, background electrolyte = 0.3 M NaCl). All solutions and suspensions were prepared using degassed Milli-Q (18 MΩ) water. The mineral formed instantly after mixing the solutions (1 – 10 ms)[4]. Mackinawite samples allowed to react for 5 mins were then drop-casted onto an Au plate for XAS (resp. to a Si plate for PES) and sealed in a glass vial inside the glove box before loading into the spectroscopy chamber via a loadlock under a $N_2$ atmosphere on the same day of sample synthesis. The short reaction time was chosen to prevent precipitation of a mixed Fe(II)/Fe(III)-S phase (e.g., greigite), which is saturated under these experimental conditions.

The XAS and PES measurements were performed on beamlines 8.0.1 and 9.3.2, respectively, at the Advanced Light Source, Lawrence Berkeley National Laboratory. For XAS, the undulator and spherical grating monochromator supply a linearly-polarized photon beam. The XAS spectra were collected by measuring both sample current (i.e., total electron yield, TEY) and fluorescent yield (i.e., total fluorescence yield, TFY). All XAS spectra were normalized to the beam flux measured with a clean gold mesh. Resolution was better than 0.2 eV, and the base pressure at RT was $8 \times 10^{-10}$ Torr during XAS measurements. The PES data were collected at a photon energy of 490 eV using a Scienta R4000 HiPP spectrometer[47]. Overall resolution was about 0.2 eV. The base pressure at RT during the PES measurements was $3 \times 10^{-10}$ Torr. We also tested air-exposed FeS samples as well as samples reacted for 65 mins in the mixed solution: both cases exhibit extra features on the main absorption peak in Fe $L$-edge XAS spectra indicating oxidation of Fe(II), which were not detected in the spectra utilized in this work. The consistency in the XAS spectra between TEY and TFY further indicated that there was in fact no significant surface oxidization.

## B. Computational details

All calculations for the NM, FM, AFM-checkerboard, AFM-single-stripe and AFM-double-stripe orderings (Fig. 2) were performed with the CASTEP code,[48] which implements DFT using a plane-wave basis set under periodic boundary conditions and ultrasoft pseudopotentials under GGA-PBE.[49] In layer-type tetrahedral Fe materials, an Fe pseudopotential treating the Fe $p$ states in the valence (as opposed to core) under PBE is essential to obtain the same stable magnetic ordering (i.e., AFM-single-stripe) as found by full-potential all-electron methods.[27] Mazin et al.[27] have pointed out that an Fe pseudopotential parameterized with the $p$ states in the core (i.e., not in the valence) will incorrectly predict the AFM-checkerboard magnetic ordering as being unstable compared to the NM ground state. In our study, we employed an Fe pseudopotential treating both $3s$ and $3p$ states in the valence, the valence electron configuration thus being *$3s^2 3p^6 3d^6 4s^{1.75}$* (for S, it was *$3s^2 3p^4$*). The number of non-local projectors was one for each $s$ state ($3s$ and $4s$) and two each for the $3p$ and $3d$ states. The core radius of Fe was 2.2 $a_0$ and that for S was 1.70 $a_0$; the value of $q_c$ for kinetic energy optimization was set at 5.5 $Ry^{1/2}$. Our pseudopotentials tested very well against published results (e.g., structural parameters and band structure) of all-electron methods for $FeS_2$ (NM ground state) and FeO (AFM ground state).[50, 51]

The plane-wave basis sets were expanded to a kinetic energy of 400 eV with a 14 × 14 × 11 k-point grid[52] or equivalent k-spacing for AFM-stripes to achieve an atomic force convergence << 0.01 eV/Å and a total energy convergence << 0.0001 eV. For the energy vs. lattice parameter curves, a 7 × 7 × 5 k-point grid or equivalent k-spacing for AFM-stripe orderings was used with the force and total energy convergence << 0.03 eV/Å and << 0.005 eV, respectively. The electronic energy was minimized with the density-mixing approach[53] using a partial occupation scheme (Gaussian smearing = 0.01 eV, with smearing effects on total energy << 0.0001 eV). The BFGS procedure[54] was followed for geometry optimization with the maximum atomic force tolerance at 0.03 eV/Å. Under DFT/GGA a clear binding energy between $FeS_4$ tetrahedral sheets is not found with respect to relaxation of the $c$ lattice parameter because of unaccounted-for van der Waals interactions between the sheets; thus the value of $c$ was fixed at 5.033 Å, the experimental value, in all geometry optimizations. When the lattice parameter $a$ (= $b$) was relaxed during geometry optimization, the residual atomic force and the root-mean-square stress after the optimization were << 0.01 eV/Å and << 0.02 GPa, respectively.

Elasticity coefficients ($C_{ij}$) were calculated for the NM and AFM-single-stripe orderings using the finite-strain technique, which determines stress tensors for a strain homogenously applied to an equilibrium structure.[55] The strain applied to the unit cell was < 0.004, and the corresponding stress was calculated implementing a 0.006 eV/Å force convergence and a 14 × 14 × 11 (or equivalent for AFM-single-stripe) k-point grid. The calculated elasticity coefficients all satisfied the Born criteria for tetragonal crystals: $C_{11} > 0$, $C_{33} > 0$, $C_{44} > 0$, $C_{66} > 0$, $C_{11} - C_{12} > 0$, $C_{11} + C_{33} - 2C_{13} > 0$ and $2(C_{11} + C_{12}) + C_{33} + 4C_{13} > 0$.[56]

### III. RESULTS AND DISCUSSION

#### A. X-ray absorption and photoelectron spectra

The Fe $L_{23}$-edge XAS spectrum corresponds to the excitation of Fe $2p$ core electrons to unoccupied Fe $3d$ states and thus is very sensitive to the local bonding environment of Fe. The mackinawite Fe $L_{23}$ XAS spectrum (Fig. 3a) shows its line shape is very similar to that for the Fe-based superconductors[57], without the multiplet structure which appears in spectra of Fe oxides such as hematite ($\alpha$-$Fe_2O_3$). While layer type FeAs compounds show a weak shoulder on the $L_3$ peak around 707 eV, mackinawite has an $L_3$ peak as sharp as that seen for Fe metal (See supplementary material, Fig. S1), which is indicative of a strong delocalization of Fe $3d$ electrons as expected from the short Fe-Fe distance (2.48, 2.60, 2.67, and 2.85 Å for Fe metal, FeS, FeSe, and $LaO_{1-x}F_xFeAs$, respectively). This narrow peak, observed in both TEY and TFY modes, confirms also that there was no discernible oxidization of the mackinawite surface during sample preparation and that the sample had essentially one type of local Fe bonding environment.

Figure 3b is the Fe $3s$ core-level PES spectrum. The spectrum can be fitted by using two clear peaks with a splitting energy of 2.9 ($\pm$ 0.2) eV. This Fe-$3s$ exchange splitting energy ($\Delta E_{3S}$) matches that of the superconductor[36], $CeFeAsF_{0.11}O_{0.89}$ ($T_c$ = 55 K), which corresponds to an Fe local magnetic moment of about 1 $\mu_B$ based on an empirical relationship between $\Delta E_{3S}$ and the spin moment in Fe-based materials (Fig. 4). [As Bondino et al.[36] (see also the references therein) have discussed in detail, the linear relationship is valid among Fe-based ionic compounds and itinerant systems, but a poor relationship is expected if a ligand bound to Fe is of low electronegativity in an insulating compound because the effects of charge transfer (from the ligand to Fe-$3d$ state) on photoemission final-state screening becomes main features in the $3s$ spectrum rather than the exchange.[58, 59]] While Mössbauer spectroscopy does not detect magnetic ordering (or the presence of an Fe magnetic moment) in mackinawite,[17, 18] the short intrinsic timescale of PES ($10^{-15}$ s, compared to $10^{-7}$ s for Mössbauer spectroscopy) allows this technique to reveal the existence of a local Fe moment in mackinawite, apparently for the first time, providing direct evidence for strong Fe-based spin fluctuations.[36]

## B. DFT optimizations

Our DFT results (Fig. 5) show that an AFM-single-stripe ground state of mackinawite is energetically more favorable than NM by 90 meV and AFM-checkerboard by 51 meV. This finding is consistent with other DFT studies of layer-type Fe-based materials (e.g., FeSe and LaO$_{1-x}$F$_x$FeAs).[27, 28, 30] Although AFM-double-stripe ordering is known to be the ground state of FeTe[60, 61] which is also isostructural with mackinawite, we found that in mackinawite double-stripe ordering is not energetically more favorable than single-stripe ordering. The FM ordering has a much higher total energy and a larger *a* lattice parameter at the energy minimum than do the other magnetic orderings. We found in fact that an initial FM ordering spontaneously transforms into a zero net magnetic moment configuration during geometry optimization if the lattice parameter *a* is lower than about 3.8 Å, as also has been reported in DFT studies of FeS,[62] FeSe,[30] and LaOFeAs.[63]

The Fe 3*s* core-level PES and DFT results both indicate a non-vanishing Fe magnetic moment in mackinawite, but the DFT estimate of the Fe magnetic moment (2.7 $\mu_B$, Table I) in the optimal AFM-single-stripe ordering is much too high as compared to the estimate made from $\Delta E_{3S}$ measured in our PES experiment. This over-estimation is another indication of itinerant spin fluctuations, as in the Fe-based superconductors.[32, 33] We have also estimated the Fe magnetic moment from a plot of the calculated exchange energy for Fe 3*s* states against the experimental spin state (Fig. 4), which shows a linear correlation. (The calculated energies appear to be systematically lower than experimental values. This can be attributed to the neglect of excited states that occur in a PES experiment.) Based on the linear correlation, we can estimate the Fe magnetic moment to be 1.6 $\mu_B$.

Our calculated density of states (DOS) for both NM and AFM-stripe FeS (Fig. 6) indicates a metallic nature, with the dominant contribution near the Fermi energy coming from Fe 3*d* states as a result of direct intralayer Fe-Fe interactions, a result which is consistent with the general DOS features of layer-type tetragonal Fe compounds.[29, 30, 64] The bandwidth at the Fermi energy is about 4.5 eV (from − 2.5 eV to 2.0 eV), which is close to that in *bcc* Fe (6.0 eV) and indicative of significant electron delocalization, but with the DOS at the Fermi energy ($E_F$) being much lower than that in *bcc* Fe, as noted also for FeSe and layer type FeAs materials.[21] In order for a FM state to be stable relative to a non-spin-polarized NM state, the Stoner criterion [$N(E_F)I > 1$] must be met,[65] where $N(E_F)$ is the DOS at the $E_F$ per atom per spin and *I* is the effective exchange interaction per pair of 3*d* electrons (typically about 0.7 to 0.9 eV for Fe). The instability of FM FeS below *a* = 3.8 Å (Fig. 5) is thus attributed to its low DOS at the $E_F$.[62] We note in passing that the contribution of S 2*p* states to the DOS is very minor near the Fermi energy, which is in fact reflected by the sharp Fe-$L_3$ peak in the XAS spectrum (Fig. 3a).

The structural parameters of geometry-optimized mackinawite are very sensitive to the type of magnetic ordering (Table I). The lattice parameter *a* and the Fe-Fe distance for NM FeS are in very good agreement with experiment, while the AFM orderings exhibit a slightly larger *a* parameter than the experimental value [by about 2 % (AFM-checkerboard) or 4 % (AFM-single-stripe)]. The Fe-S distance in NM FeS is smaller by about 3 % than the experimental value, while those for the AFM variants are very close to experiment (< 1 % over-estimation). In the double-stripe ordering, *d*(Fe-Fe) between Fe with parallel spins was shorter than that between Fe with anti-parallel spins, leading to P2$_1$/m symmetry within the tetragonal lattice. Full relaxation of the structure of AFM-double-stripe mackinawite turned it into a non-magnetic state. Our elasticity coefficient ($C_{ij}$) calculations show a large compressional elastic anisotropy for both NM and AFM FeS, with AFM FeS being the more anisotropic: $C_{11}/C_{33}$ = 5.2 for NM; $C_{11}/C_{33}$ = 6.8 for AFM-checkerboard; $C_{11}/C_{33}$ = 16.2 for AFM-single-stripe. This large anisotropy is attributed to weak van der Waals interactions among sheets stacked along the *c* axis, as opposed to the stronger bonding among the FeS$_4$ tetrahedra within a sheet. Ehm et al.[66] reported a highly anisotropic lattice compressibility based on high-pressure experiments, with the linear compressibility along the *c* axis being 5 to 7 times larger than that along the *a* axis.

The NM calculation appears to reproduce the experimental mackinawite structure well. However, the *z*-coordinate of S along the *c* axis ($Z_S$, Table I) is much more poorly predicted in NM FeS, even when constrained at the experimental lattice parameters (Table II), than that in AFM FeS, as is also seen for the As z-coordinate predicted in FeAs compounds.[27, 33, 46] Magnetic ordering yields a much better $Z_S$; in particular, excellent agreement with experiment is achieved with the AFM-single-stripe ordering with the experimental lattice parameters (Table II). We also found a sensitive relationship between structural parameters and the local Fe magnetic moment (Fig. 7). The latter drops rapidly with decreasing lattice *a* parameter and, therefore, Fe-S distance and $Z_S$, particularly if *a* is near its experimental value (3.7 Å). This distinctive dependence of the magnetic moment on structural parameters, also observed in FeAs-based superconductors, is yet another indicator of strong itinerant spin fluctuations.[32, 33, 46]

Our Fe 3*s* core-level PES and DFT results both indicate the existence of strong itinerant spin fluctuations in mackinawite. We therefore conclude that mackinawite exhibits the magnetic characteristics of the high temperature Fe-based superconductors with which it is isostructural. If strong spin fluctuations are the mediators of electron pairing in these Fe-based superconductors, our findings allow us to conjecture that mackinawite may be one of the simplest Fe-based superconductors.

## ACKNOWLEDGMENTS

Our computations used resources of the National Energy Research Scientific Computing Center, which is supported by the Office of Science of the U.S. Department of Energy under Contract No. DE-AC02-05CH11231. K.D.K. thanks Dr. A. Mehta for helpful discussion on experimental approaches. This research reported in this paper was supported by the Director, Office of Energy Research, Office of Basic Energy Sciences, of the U.S. Department of Energy under Contract No. DE-AC02-05CH11231.

## Supplementary materials (EPAPS)

Fe *L*-edge XAS spectra of Fe, FeS, $SmO_{0.85}FeAs$, and $Fe_2O_3$ (Fig. S1).

FIG. 1. (Color online) Crystal structure of mackinawite. Blue = Fe; Orange = S.

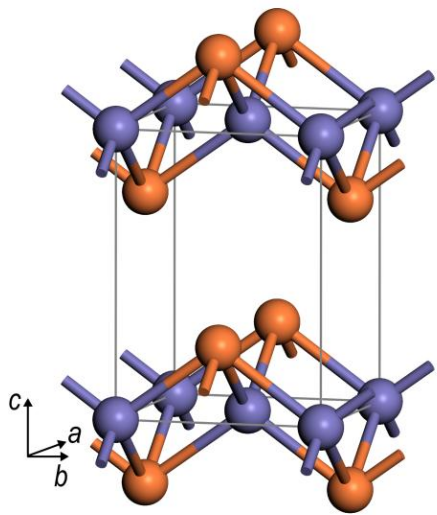

FIG. 2. (Color online) Three possible antiferromagnetic (AFM) orderings in mackinawite: (a) AFM-checkerboard, (b) AFM-single-stripe, and (c) AFM-double-stripe. Purple = spin-up Fe; Yellow = spin-down Fe. Solid line represents a crystallographic unit cell and dotted line represents a magnetic unit cell for each ordering. The Fe spin moments are ferromagnetically-aligned between sheets.

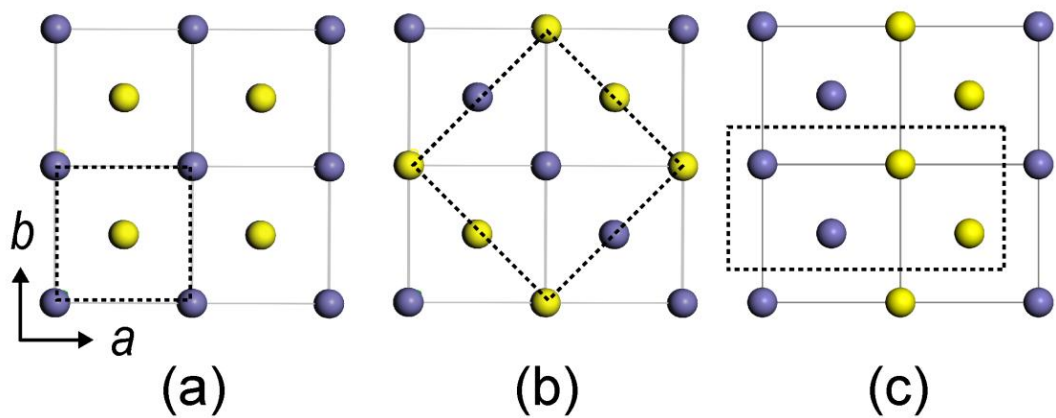

FIG. 3. Experimental (a) Fe $L_{23}$ XAS spectra and (b) Fe $3s$ core-level PES spectrum. TEY: total electron yield; TFY: total fluorescence yield. $\Delta E_{3s}$: multiplet splitting.

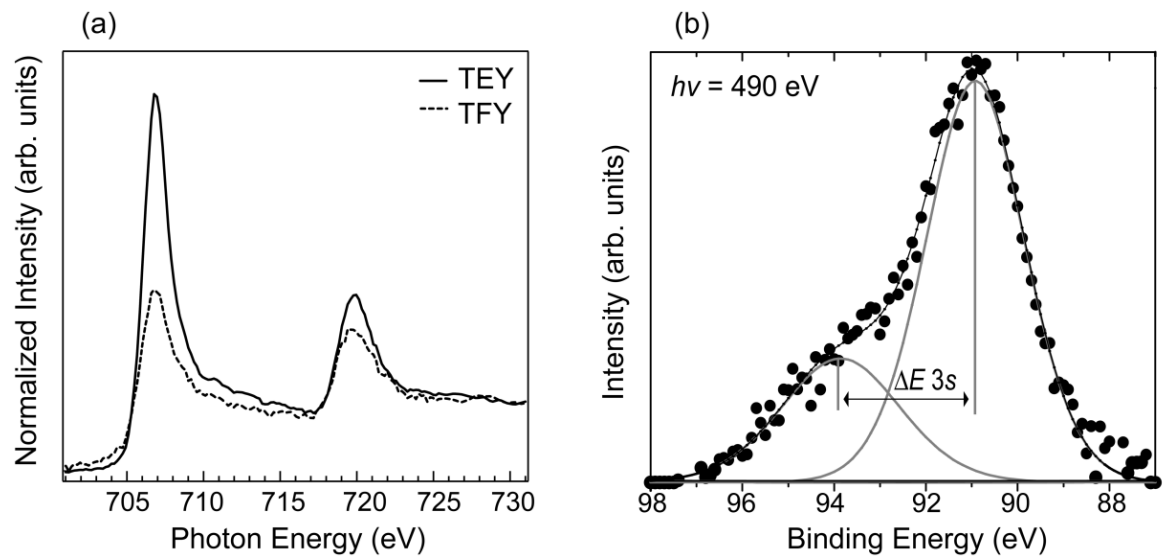

FIG. 4. (Color online) Relationship between the Fe 3$s$ multiplet splitting ($\Delta E_{3s}$) and the net spin state (magnetic moment ≈ 2S $\mu_B$, where S designates a spin state) for Fe-based materials. The experimental data (except for FeS) are from Ref. 36 and references cited therein. The calculated $\Delta E_{3s}$ is the exchange energy of the 3$s$ core-level in geometry-optimized FeS (AFM-single-stripe), Fe (FM), FeO (type II-AFM) and FeF$_3$ (G-type AFM) as obtained in the current study.

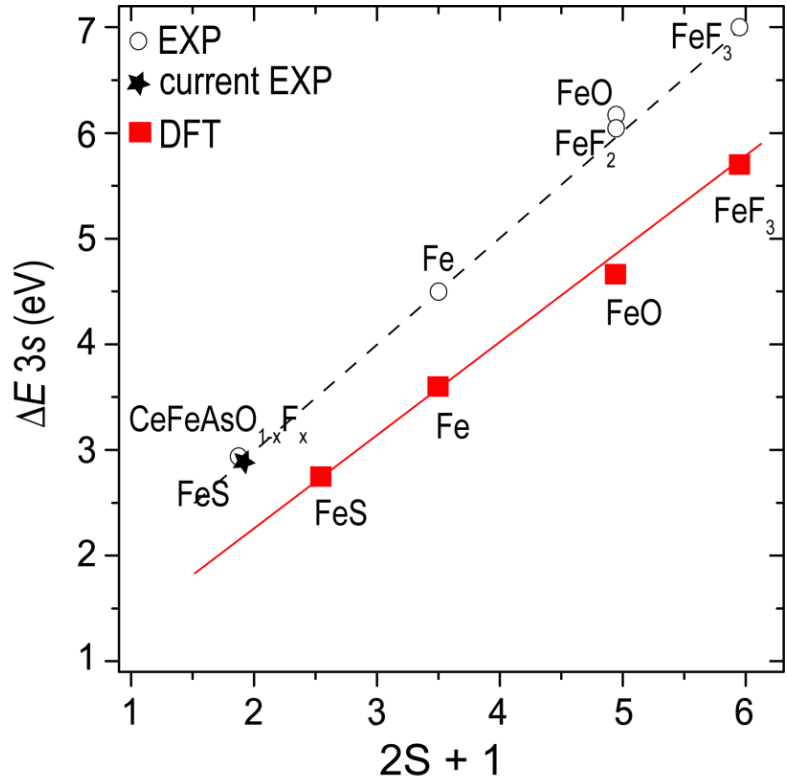

FIG. 5. (Color online) Total energy per formula unit vs. the lattice parameter *a* in mackinawite for different magnetic orderings. Vertical line denotes the experimental *a* value.

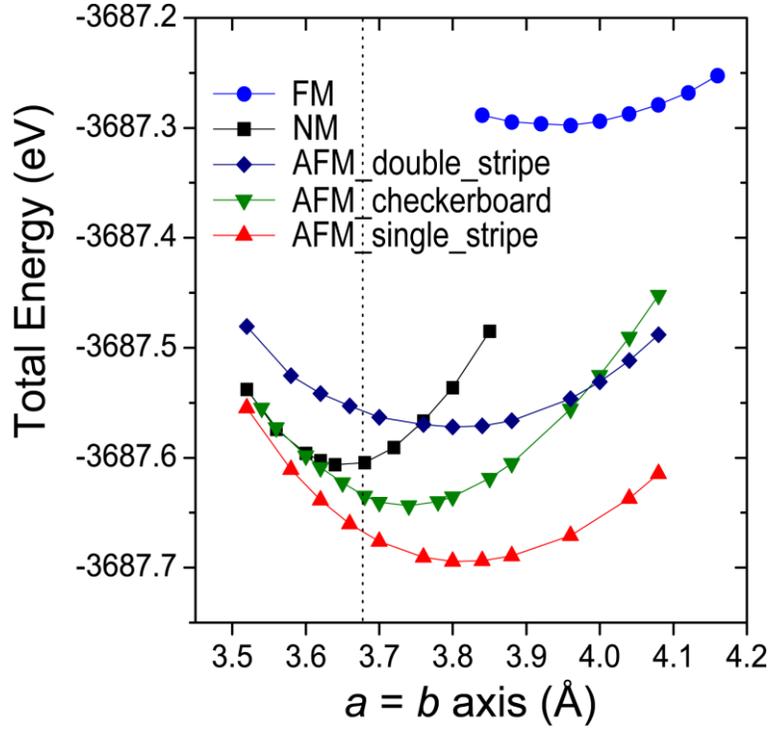

FIG. 6. (Color online) Partial density of states (DOS) of geometry-optimized mackinawite in (a) NM ($a$ = 3.655 Å) and (b) AFM-single-stripe ($a$ = 3.832 Å) magnetic ordering. Vertical line represents the Fermi energy.

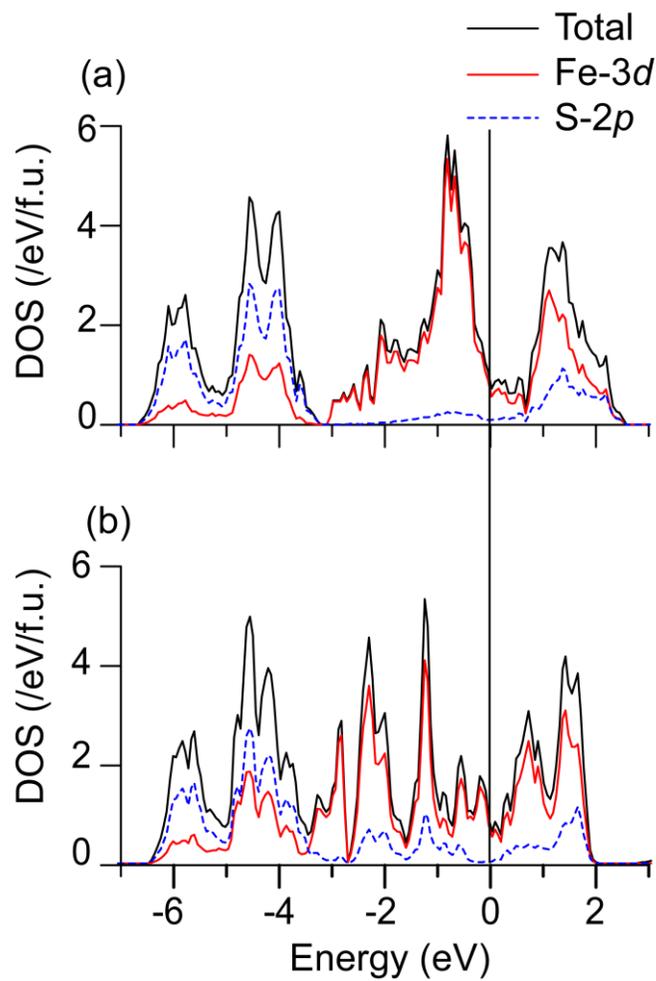

FIG. 7. Dependence of the calculated Fe magnetic moment of AFM-single-stripe mackinawite on the lattice parameter *a*. Vertical line represents the experimental *a* (= *b*) value.

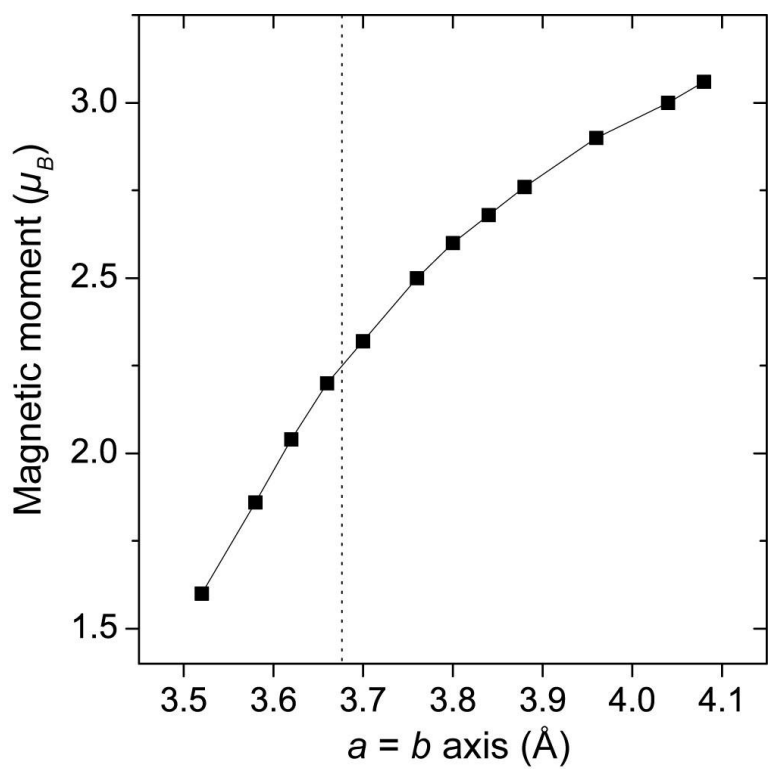

TABLE I. Results of geometry optimization of mackinawite in different magnetic orderings[a].

| | NM | AFM | | | FM | experiment[b] |
| --- | --- | --- | --- | --- | --- | --- |
| | | checker board | single-stripe | double-stripe | | |
| $a$ (Å) | 3.655 | 3.752 | 3.833 | 3.814 | 3.961 | 3.674 |
| $d$(Fe-S) (Å) | 2.192 | 2.258 | 2.287 | 2.347[c] | 2.378 | 2.256 |
| $d$(Fe-Fe) (Å) | 2.584 | 2.653 | 2.710 | 2.659/2.736 | 2.801 | 2.598 |
| $z_S$[d] | 0.240 (-0.10 Å) | 0.250 (-0.05 Å) | 0.248 (-0.06 Å) | 0.258 (-0.01 Å) | 0.262 (+0.01 Å) | 0.260 |
| Fe moment ($|\mu_B|$) | 0.0 | 2.4 | 2.7 | 3.0 | 3.4 | |
| Total energy[e] (meV) | 0 | -39 | -90 | +32 | +308 | |

[a] NM: non-magnetic; AFM: antiferromagnetic; FM: ferromagnetic ordering.
[b] Ref. 2.
[c] Additional $d$(Fe-S): 2.264 Å and 2.351 Å. See text.
[d] The z-coordinate of S along the $c$-axis. Value in parenthesis is the deviation of S position from experiment.
[e] Total energy per formula unit relative to that of NM.

TABLE II. Structure, Fe moment, and total energy of mackinawite under DFT internal geometry optimization constrained by the experimental lattice parameters[a].

|  | NM | AFM checkerboard | AFM single-stripe | experiment[a] |
|---|---|---|---|---|
| $d$(Fe-S) (Å) | 2.195 | 2.238 | 2.248 | 2.256 |
| $Z_S$ | 0.239 | 0.254 | 0.257 | 0.260 |
| Fe moment ($\mu_B$) | 0 | 2.2 | 2.2 | |
| Total energy[b] (meV) | 0 | -27 | − 61 | |

[a] Experimental data from Ref. 2. Coordinates of all atoms were relaxed during geometry optimization with the lattice parameters fixed at $a = b = 3.674$ Å and $c = 5.033$ Å.

FIG. S1. Comparison of Fe *L*-edge XAS spectra for Fe metal, mackinawite, $SmO_{0.85}FeAs$, and $Fe_2O_3$, as measured in the beamline 8.0 in the Advanced Light Source. Dashed curve, TFY mode; all others are TEY mode. The spectrum of Fe metal is from sputtered iron film. Details of the experimental conditions for Fe, $SmO_{0.85}FeAs$, and $Fe_2O_3$ are described elsewhere [W. L. Yang, et al., Phys. Rev. B **80**, 014508 (2009)].

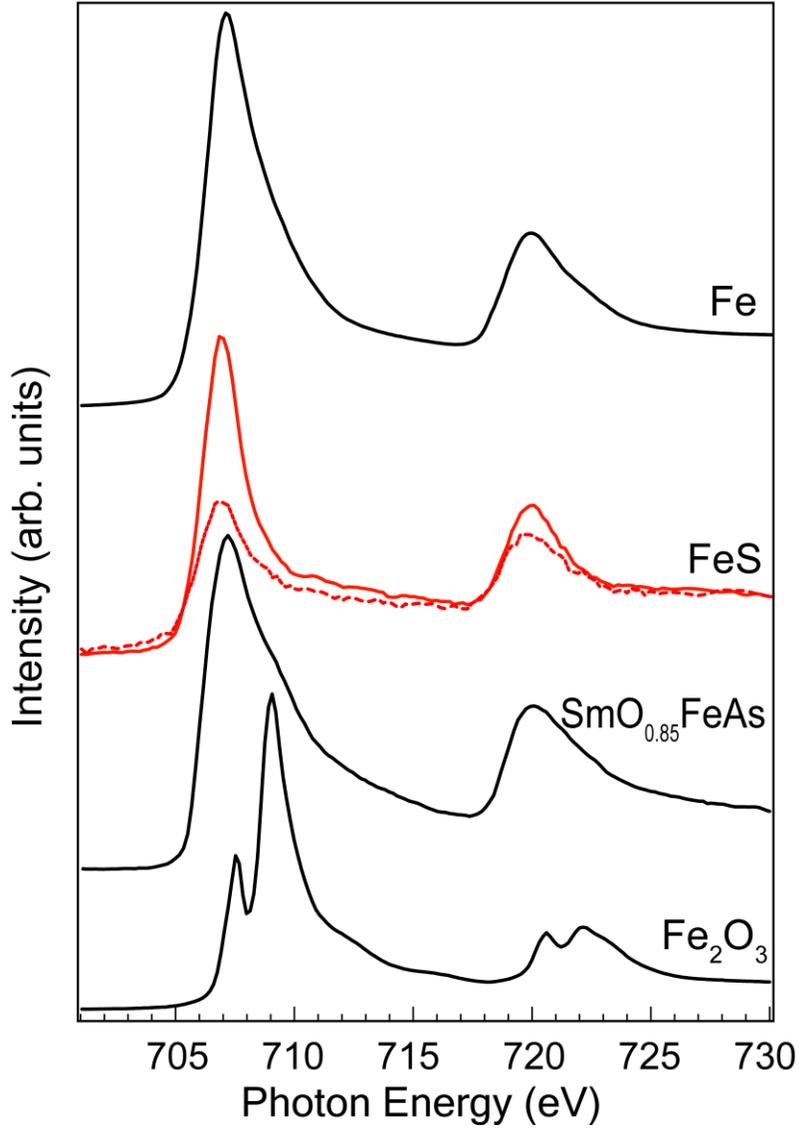